\documentclass[preprint]{aastex}
\usepackage{natbib}
\usepackage{booktabs}
\usepackage{threeparttable}
\usepackage{CJK}

\begin{document}
\title{Quasi-periodic pulsations of $\gamma$-ray emissions from a solar flare on 2017 September 06}

\author{D.~Li\altaffilmark{1,2}, D.~Y.~Kolotkov\altaffilmark{3,4}, V.~M.~Nakariakov\altaffilmark{3,5}, L.~Lu\altaffilmark{1}, and Z.~J.~Ning\altaffilmark{1}}
\affil{$^1$Key Laboratory of Dark Matter and Space Astronomy, Purple Mountain Observatory, CAS, Nanjing 210033, China \\
    $^2$State Key Laboratory of Space Weather, Chinese Academy of Sciences, Beijing 100190, China \\
    $^3$Centre for Fusion, Space and Astrophysics, Department of Physics, University of Warwick, Coventry CV4 7AL, UK \\
    $^4$Institute of Solar-Terrestrial Physics, Lermontov St., 126a, Irkutsk 664033, Russia \\
    $^5$St. Petersburg Branch, Special Astrophysical Observatory, Russian Academy of Sciences, 196140, St. Petersburg, Russia
    }
\altaffiltext{2}{Correspondence should be sent to:
lidong@pmo.ac.cn.}

\begin{abstract}
We investigate quasi-periodic pulsations (QPPs) of high-energy
nonthermal emissions from an X9.3 flare (SOL2017-Sep-06T11:53), the
most powerful flare since the beginning of solar cycle 24. The QPPs
are identified as a series of regular and repeating peaks in the
light curves in the $\gamma$- and hard X-ray (HXR) channels recorded
by the {\it Konus-Wind}, as well as the radio and microwave fluxes
measured by the CALLISTO radio spectrograph during the impulsive
phase. The periods are determined from the global wavelet and
Fourier power spectra, as 24$-$30~s in the HXR and microwave
channels which are associated with nonthermal electrons, and
$\sim$20~s in the $\gamma$-ray band related to nonthermal ions. Both
nonthermal electrons and ions may be accelerated by repetitive
magnetic reconnection during the impulsive phase. However, we could
not rule out other mechanisms such as the MHD oscillation in a
sausage mode. The QPP detected in this study is useful for
understanding the particle acceleration and dynamic process in solar
flares and also bridging the gap between stellar and solar flares
since the energy realm of the X9.3 solar flare is almost compared
with a typical stellar flare.
\end{abstract}


\section{Introduction}
Quasi-periodic pulsations (QPPs) are a common feature in flaring
emissions on the Sun and Sun-like stars. They are frequently
detected as regular periodic peaks in light curves of solar/stellar
flares, based on the time-series analysis \citep[e.g.,][and
references therein]{Nakariakov09,Pugh16,Van16,McLaughlin18}. The
flare-related QPPs can be observed in almost all the electromagnetic
wavebands, i.e., radio
\citep{Aschwanden94,Kupriyanova16,Nakariakov18}, H$\alpha$
\citep[e.g.,][]{Srivastava08}, L$\alpha$
\citep[e.g.,][]{Milligan17}, ultraviolet/extreme-ultraviolet
\citep[UV/EUV,][]{Nakariakov99,Kumar16,Li16}, soft/hard X-rays
\citep[SXR/HXR,][]{Zimovets10,Tan16,Hayes19}, and even $\gamma$-rays
\citep{Nakariakov10}. Furthermore, QPPs are studied in spectroscopic
observations, such as the Doppler velocity, line intensity and width
of hot emission lines \citep[e.g.,][]{Wang02,Tian16,Brosius18}. The
characteristic periods of flare-related QPPs appear in a broad
range, i.e., from sub-seconds through seconds to dozens of minutes
\citep{Karlicky05,Shen12,Inglis16,Kolotkov18}. Often, QPPs show a
similar period in a broad wavelength range
\citep{Dolla12,Kumar16,Ning17}. On the other hand, the QPPs in a
same event can exhibit multiple periods in a single waveband
\citep{Chowdhury15,Kolotkov15,Li17}. In some cases, the period
ratios are found to correspond to the period ratios typical for
magnetohydrodynamic (MHD) modes, e.g., sausage modes
\citep{Inglis09}.

Up to now, the physical mechanism responsible for the generation of
QPPs is still an open issue \citep[see,][for the reviews of various
theoretical models]{Van16,McLaughlin18}. In particular,
flare-related QPPs observed in nonthermal emissions (i.e., radio,
microwave, HXR and $\gamma$-rays) are associated with accelerated
electrons or ions, which might be produced by a periodic energy
release \citep{Kliem00,McLaughlin18}. The short-period QPPs detected
in radio/microwave emissions may be related to the dynamic
interaction between waves and energetic particles, while the
long-period QPPs observed in white light, UV and EUV wavebands are
usually thought to be associated with the dynamics of the emitting
plasmas \citep{Aschwanden87,Nakariakov06,Nakariakov09}. In
particular, QPPs can be driven by periodically induced magnetic
reconnection (e.g., by MHD oscillations), or may be modulated by MHD
waves such as slow, kink, and sausage waves, or could be a signature
of spontaneous repetitive magnetic reconnection (i.e., by magnetic
dripping mechanism). It is likely that different classes of QPPs are
produced by different mechanisms
\citep{Nakariakov09,Van16,McLaughlin18}.

QPPs in the $\gamma$-ray band are rarely reported. To date, only a
40-s QPP was found in $\gamma$-ray flux in an X1.7 flare
\citep{Nakariakov10}. In this study we analyze the X9.3 flare on
2017 September 06 which has been the most powerful flare since 2005.
Its released energy could even be in the realm of a typical stellar
flare \citep[see,][]{Kolotkov18}. In this paper, we demonstrate the
presence of QPPs in radio, microwave, HXR and also $\gamma$-ray
channels in this flare. Our observational results could be helpful
for understanding the particle acceleration and dynamic process in
most powerful solar/stellar flares
\citep{Nakariakov10,McLaughlin18}.

\section{Observations}
On 2017 September 06, the most powerful flare (X9.3) of solar cycle
24 occurred in the active region of NOAA~12673. It started at
$\sim$11:53~UT, peaked at around 12:02~UT, and ended at
$\sim$12:10~UT in {\it GOES} SXR light curves at 1$-$8~{\AA}, as
shown by the black curve in Figure~\ref{image}~(a), a short vertical
line indicates the flaring peak time. The X9.3 flare was recorded by
the {\it Konus-Wind} instrument during the impulsive phase in the
HXR and $\gamma$-ray channels, such as G1~22$-$83~keV,
G2~83$-$331~keV, and G3~331$-$1253~keV, as can be seen by the color
lines in panel~(b), here the time range is outlined by two magenta
arrows in panel~(a). {\it Konus-Wind} is a US-Russian experiment
which aims to investigate the solar flares and $\gamma$-ray bursts.
It operates in two modes: the daily waiting mode with accumulation
time of 2.944~s, and the triggered mode with a nonuniform time
cadence varying from 2~ms to 256~ms with a total duration of
$\sim$250 s \citep{Aptekar95,Palshin14}. Noting that the 3-s
periodic dips with a duration of 30~ms in light curves are caused by
instrumental effects due to the {\it Konus-Wind}
occultation\footnote{http://www.ioffe.ru/LEA/kwsun/}. Therefore, the
light curves measured from {\it Konus-Wind} are firstly interpolated
to a uniform time cadence of 1.024~s, so the 3-s periodic dips are
inapparent in Figure~\ref{image}~(b).

Figure~\ref{image}~(c)$-$(e) show the HMI continuum filtergram,
AIA~1600~{\AA} image, and HMI line-of-sight (LOS) magnetogram with
the same field of view of around 150\arcsec~$\times$~150\arcsec,
respectively. They have been pre-processed by `aia\_prep.pro' and
`hmi\_prep.pro' in solar soft ware \citep{Lemen12,Schou12}. It can
be seen that the powerful flare occurs in the active region which is
apparently associated with a $\delta$-type sunspot, as shown in
panel~(c). The X9.3 flare emits strong UV emission at the center
region nearby (x=530\arcsec, y=$-$250\arcsec), but it emits weak UV
radiation at the north (such as x=580\arcsec, y=$-$200\arcsec) and
south (i.e., x=560\arcsec, y=$-$290\arcsec) regions. The strong UV
radiation is underlying a strong EUV emission observed in
AIA~131~{\AA} (magenta contours) and 94~{\AA} (cyan contours)
wavelengths, as shown in panel~(d). The strong EUV emission could be
attributed to hot flaring loops linking the apparent flare ribbons.
On the other hand, the strong UV radiation is overlaying a very
complex and sheared magnetic polarity regions, which are composed of
several strong positive and negative polarity sources, as indicated
by the red and green contours in the center of panel~(e). Based on
the overall distribution of the positive and negative fields, the
X9.3 flare is most likely a double-ribbon flare connected by hot
flaring loops. The observed flare morphology is consistent with the
standard `CSHKP' model
\citep{Carmichael64,Sturrock66,Hirayama74,Kopp76}, or 2-D
reconnection model \citep{Sturrock64}. According to these models,
the plasma at the reconnection region could be heated to more than
10 MK, and electrons will be efficiently accelerated to nonthermal
energies. Subsequently, the released energy will be transported away
from the reconnection site via nonthermal particles towards
footpoints of the reconnection loop, and also outward along the open
magnetic field. In this process, the flaring loop can be clearly
seen in SXR or EUV wavelength, with the HXR or microwave sources
situated near the footpoints or loop top, and the flaring ribbons
formed in the visible or UV waveband \citep[e.g.,][and references
therein]{Masuda94,Lin05,Fletcher11,Benz17,Yan18}. Moreover, the type
III radio bursts are often accompanied by solar flares, which are
thought to be the signatures of propagating beams of nonthermal
electrons in the solar corona \citep[e.g.,][]{Reid14}. In our study,
the flaring ribbons seem to be fragmented in UV~1600~{\AA} image due
to the image saturation of AIA observations during flare eruptions
\citep[see discussion of this issue in][]{Lemen12}.

The X9.3 flare was also measured by the ground-based CALLISTO radio
spectrograph \citep{Benz09} located at GREENLAND (a) and BLEN5M (b),
as shown in Figure~\ref{spect}. Panel~(a) presents the radio dynamic
spectrum at lower frequencies from $\sim$16.8~MHz to
$\sim$100.6~MHz, and the over-plotted light curve is the radio flux
at a frequency of around 60.8~MHz, as indicated by the short cyan
line on the left-hand. A series of transient bursts can be seen in
the radio dynamic spectrum during the impulsive phase, i.e., between
around 11:57~UT and 12:00~UT. They are characterized by a short time
duration and a fast frequency drift, i.e., drifting quickly from
higher to lower frequencies over a short time. The observed behavior
is consistent with typical type III radio bursts, which can be used
for tracing the propagating beams of flare-accelerated nonthermal
electrons through the solar atmosphere
\citep[e.g.,][]{Wild50,Reid14}. Meanwhile, much stronger bursts
appear after the peak time of the flare, such as $\sim$12:02~UT. The
strong bursts exhibit a slow downward frequency drift, and the
fundamental and harmonic frequencies are clearly seen in the radio
dynamic spectrum. Thus, the slow drifts can be regarded as type II
radio bursts, which are triggered by the electron beams accelerated
by shock waves \citep[e.g.,][]{Wild50,Makela18}. However, as {\it
Konus-Wind} measured the HXR and $\gamma$-ray emission in the
flaring impulsive phase, i.e., before $\sim$12:00~UT, in this paper
we analyse only the fast drifting radio bursts which could be
directly compared with the HXR and $\gamma$-ray emission. Panel~(b)
gives the radio spectrogram at higher frequencies between
$\sim$1040~MHz and $\sim$1436~MHz, generally considered as the
microwave radiation. Similar to the radio dynamic spectrum at lower
frequencies, the microwave spectrogram is also dominated by two
pieces of strong emissions, one is characterized by a fast frequency
drift between about 11:57~UT and 12:00~UT during the impulsive
phase, and the other one exhibits a slow frequency drift after
12:00~UT. The over-plotted light curve is the microwave flux at a
frequency of nearby 1250.9~MHz. Both the radio/microwave fluxes at
the frequencies of 60.8~MHz and 1250.9~MHz are interpolated to a
uniform time cadence of 1.0~s, which is close to that of the {\it
Konus-Wind} light curves.

\section{Data Analysis and Result}
All the flaring fluxes during the impulsive phase recorded by {\it
Konus-Wind} and CALLISTO appear to exhibit a signature of QPPs,
i.e., a series of regular and repeating pulsations with a period of
roughly 30~s in the light curves, as can be seen in
Figures~\ref{image}~(b) and \ref{spect}~(a). To study these
flare-related QPPs in detail, we perform wavelet analysis on the
detrended light curves after removing a 30-s running average
\citep[e.g.,][]{Yuan11,Tian16,Li18}, as shown by the black dashed
lines in Figures~\ref{image}~(b) and \ref{spect}. The detrended
light curves are used because we thereby enhance the shorter-period
oscillations and suppress the long-period trend \citep[see][for the
discussion and justification of this
method]{Gruber11,Kupriyanova10,Kupriyanova13,Auchere16}. Here, the
wavelet power is normalized in accordance with the Parseval's
theorem for wavelet analysis \citep[see,][]{Torrence98}, providing
conservation of the total energy of the signal under the wavelet
transform and thus allowing us to obtain distribution of the
spectral power across wavelet periods.

Figure~\ref{wave1}~(a) shows the detrended and normalized to the
maximum fluxes measured by {\it Konus-Wind} in the HXR (G1 \& G2)
and $\gamma$-ray (G3) channels during the impulsive phase, i.e., in
the time interval of $\sim$11:54:00$-$11:59:11~UT. They all appear
to have periodic patterns, and the signals in the G1$-$G3 channels
look approximately co-phased. But the onset time in the G1 channel
seems to be earlier than that in the G2 channel. Panels~(b)$-$(d)
present wavelet analysis results from the detrended light curves,
all of which exhibit a coexistence of multiple periods, i.e., a long
one about 70~s, a short one around 30~s, and a much shorter period
nearby 10~s. The wavelet analysis spectra show that the onset time
of QPPs in the G1 channel is earlier than that in the G2 \& G3
channels, and their duration times shorten with the emission energy
from HXR to $\gamma$-ray channels. From the confidence levels at
99.9\% (see the large-region red contours which contain the bright
region), we can see that the QPPs with a period of $\sim$30~s in G1
channel start at around 11:55:16~UT and keep roughly 120~s, while
the similar QPPs in G2 channel begin at about 11:55:37~UT and remain
for around 110~s, and the QPPs in G3 channel appear at nearby
11:55:33~UT and only last for about 75~s. Panels~(e)$-$(g) present
global wavelet results in three channels, which clearly show the
presence of QPPs with a period of nearby 30~s. On the other hand,
the spectral peak corresponding to the period of about 70~s is much
lower than the confidence levels (red line), which could be
attributed to the leakage of a slowly varying component of the
background flaring trend and this periodicity is not considered in
this paper. Here, the dominant period is identified as the peak
value in the global wavelet power spectrum, and its error bar is
determined as the half full-width-at-half-maximum of the peak global
power \citep[as performed by][]{Li18}. Thus, the periods in the
G1$-$G3 channels are estimated to be $\sim$30$\pm$6~s,
$\sim$28$\pm$8~s, $\sim$20$\pm$6~s, respectively. The estimated mean
values of dominant periods seem to become shorter and shorter from
HXR to $\gamma$-ray channels. However, they overlap within the
estimated error bars. Finally, a much shorter period nearby 10~s
(above the confidence levels) can be seen in the G1 \& G2 channels,
but it is below the confidence level in the G3 channel, as shown in
panel~(e)$-$(g). However, these around 10-s oscillations look rather
sporadic and patchy which may indicate in favor of their noisy
origin (panels~b$-$d), and the type of noise which they may belong
to could be different from that considered in the wavelet analysis.
Therefore, they are also disregarded in our study.

Figure~\ref{wave2} presents the same analysis results of radio
emissions at low (60.8~MHz) and high (1250.9~MHz) frequencies
between $\sim$11:56:00~UT to $\sim$11:59:59~UT. Panel~(a) gives the
normalized detrended fluxes at the frequencies of 60.8~MHz and
1250.9~MHz, which exhibit a number of regular and repeating peaks.
Moreover, the peak times at these two frequencies are different,
suggesting a frequency drift between them. Panels~(b) and (c)
display the wavelet power spectra, which show a nearly same onset
time ($\sim$11:56:58~UT \& $\sim$11:57:00~UT) of the QPPs but
obviously different lifetimes ($\sim$130~s \& $\sim$85~s), as
indicated by the confidence levels of 99.9\% (red contours include
the bright regions). Panels~(d) and (e) show the global wavelet
power spectra, from which we can see that only one peak is above the
confidence level (red line) at the low and high frequencies,
respectively. So the periods with error bars can be estimated to be
$\sim$20$\pm$6~s at the frequency of 60.8~MHz, and $\sim$24$\pm$8~s
at the frequency of 1250.9~MHz.

We further perform Fourier analysis of the original (but not
detrended) light curves with the fast Fourier transform method
\citep[see,][]{Ning17}, as shown in Figure~\ref{fftp}. It can be
seen that each Fourier power is dominated by a power law, which is
usually identified as red noise in the astrophysical observations
\citep{Vaughan05}. The red noise could be described by a power-law
model, such as $P(f)~\sim~f^{\alpha}$. Here, $f$ is the frequency,
$\alpha$ represents a negative slope. Meanwhile, a flat spectrum in
the higher frequency region is often referred to as white noise.
Such a superposition of red and white noise, dominating at lower and
higher frequencies, respectively, is often observed in the solar
atmosphere \citep[e.g.,][]{Inglis15,Kolotkov16,Ning17}.
Figure~\ref{fftp}~(a) and (c) show that only one subpeak (magenta
arrow) appears above the confidence levels (green lines) in the
power-law spectra, confirming the presence of shorter periods
($\sim$30~s and $\sim$20~s) in the two channels. On the other hand,
there are two subpeaks (cyan and magenta arrows) above the
confidence level in panel~(b), suggesting the longer ($\sim$70~s)
and shorter ($\sim$28~s) periods might be statistically significant
in the G2 channel. According to Figure~\ref{image}~(b), the ~70-s
periodicity in the original light curves does not last for three
complete oscillation cycles and could be attributed to the slow
variations of the background flaring trends. As such, it might be a
signature of localized transient activities, viz., episodic
reconnection in the flare current sheet \citep[e.g.,][]{Jelinek17}
or slow magnetoacoustic oscillations \citep[e.g.,][]{Nakariakov19}.
However, it is out of the scope of this study, since it only appears
to be significant in the HXR (G2) channel, as shown in
Figure~\ref{fftp}~(b), and attributed to the long-period flare
trend. We also notice that there are subpeaks (blue arrow) in flat
spectra of all the {\it Konus-Wind} light curves. They are not
likely to represent real QPPs but could be caused by the 3-s
periodic dips in the light curves which was due to the {\it
Konus-Wind} occultation \citep{Aptekar95,Palshin14}. All these
Fourier power spectra do not show a subpeak nearby 10~s, further
supporting that the 10~s period in Figure~\ref{wave1} is not real
QPPs. Finally, panel~(d) shows that only one subpeak (magenta arrow)
is above the confidence level in the whole Fourier power spectrum,
confirming the shorter period of $\sim$20~s in the radio frequency
of 60.8~MHz.

\section{Conclusion and Discussion}
Using the multi-instrumental observations with the {\it Konus-Wind}
and the CALLISTO spectrographs (GREENLAND \& BLEN5M), QPPs of
high-energy nonthermal emissions are investigated in the X9.3 flare
on 2017 September 06. Our primary results are summarized as
following:

\begin{enumerate}
\item QPPs are found in the HXR and $\gamma$-ray
channels, and their periods are estimated as roughly 25~s, or more
specifically, as $\sim$30$\pm$6~s, $\sim$28$\pm$8~s,
$\sim$20$\pm$6~s in the energy channels of G1~22$-$83~keV,
G2~83$-$331~keV, and G3~331$-$1253~keV, respectively. They are
observed in the impulsive phase of the flare.

\item QPPs are also seen in radio/micowave emissions at
both low and high frequencies, whose periods are estimated as
$\sim$20$\pm$6~s and $\sim$24$\pm$8~s, which are consistent with the
periods detected in the $\gamma$-ray and HXR wavebands,
respectively. However, the QPP detected in the radio band are
delayed in comparison with the high-energy QPP by $\sim$90~s, i.e.,
about three cycles of the oscillation.

\item The QPPs observed in $\gamma$-ray emission are possibly
associated with the accelerated ions, while the QPPs seen in the HXR
and microwave emissions are most likely related to the accelerated
electrons. Both the accelerated ions and electrons can be produced
by a periodic regime of magnetic reconnection.
\end{enumerate}

It is interesting that the 20-s QPPs are detected in the flaring
$\gamma$-ray emission, which is thought to be related to nonthermal
ions. Indeed, a high-energy part of the spectrum of this flare has
been analyzed in detail by \cite{Lysenko19}, and they found the
contribution from nuclear de-excitation lines, i.e., the lines at
511~keV and 2223~keV, which confirms that the X9.3 flare does
produce accelerated ions. On the other hand, previous observations
have shown that the flaring flux in the $\gamma$-ray band could have
QPPs \citep{Chupp83}, and then the 40-s QPPs in the flaring
$\gamma$-ray emission were detailed studied by \cite{Nakariakov10}.
They concluded that the $\gamma$-ray QPPs were associated with
accelerated ions produced by magnetic reconnection which is
periodically modulated a global kink oscillation in a coronal loop
situated nearby \citep{Nakariakov06}. In this paper, the
$\gamma$-ray QPPs are detected during the impulsive phase of an X9.3
flare, and are most likely caused by nonthermal ions accelerated by
magnetic reconnection too. The period of $\sim$20~s might be
triggered by the self-induced regimes of repetitive reconnection, or
could be caused by MHD oscillations
\citep{Nakariakov06,Nakariakov09}. However, it is impossible to
determine the specific mechanism for the $\sim$20-s periodicity due
to the lack of the simultaneous imaging observations with the
required time resolution, e.g., shorter than several seconds.

It is worthwhile to stress that the QPPs detected in different
wavebands have slightly different periods ranging from 20 to 30 s.
The detected dominant periods shorten from HXR/microwave to
$\gamma$-ray emissions, such as 24$-$30 s in HXR and microwave
channels, but only 20~s in the $\gamma$-ray emission. Such a
difference might be attributed to the different flare regions, which
is first proposed by \cite{Nakariakov10}, who found that the HXR and
$\gamma$-ray emissions came from two different sources. However, we
cannot conclude it here, because the absence of imaging observations
in the microwave, HXR and $\gamma$-ray channels. Conversely, the
closest periods are found in HXR and microwave channels, which maybe
due to that their radiation sources are close to each other. On the
other hand, taking into account the fact that the QPP periods
detected in different wavebands coincide within the error bars, they
might have the same value of about 25~s, and the slight difference
could be ignored. More specifically, the observed similarity of the
QPP patterns in the emission associated with the nonthermal
electrons and ions indicates that either the acceleration or
kinematics of those two species are modulated by the same
quasi-periodic mechanisms. In the former case, the phenomenon of
self-induced repetitive reconnection, i.e., the magnetic dripping,
has been seen in numerical experiments
\citep[e.g.][]{Thurgood19,Liu19}. But, the relationship between the
oscillation period and parameters of the reconnecting plasma
configuration for realistic values of transport coefficients, in
particular, the Lundquist number, needs to be established. The
reconnection rate could also be modulated by an MHD oscillation
\citep[e.g.,][]{Nakariakov06}, while the efficiency of this
mechanism has not been studied yet. Moreover, oscillatory motions of
coronal plasma structures with the period of about 25~s could not be
resolved with the available imaging telescopes. In the latter case,
kinematics of the charged particles could be affected by a periodic
variation of the cross-sectional area of the magnetic flux tube
filled in by those particles, by a sausage oscillations, i.e., by
the Zaitsev--Stepanov mechanism \citep{Zaitsev82}. However, a
comparative study of this effect on electrons and ions has not been
performed yet, and, as mentioned above, the mother 25-s sausage
oscillation could not be spatially resolved. Finally, to address
these issues, we need more observations and cases, in particular the
simultaneous imaging observations in the microwave, HXR, and
$\gamma$-ray channels.

Based on the standard flare model
\citep[e.g.,][]{Sturrock64,Masuda94,Lin05}, both nonthermal
electrons and ions are produced by the magnetic recognition during
the impulsive phase of a solar flare, while the HXR and microwave
emissions of a solar flare are attributed to nonthermal electrons,
and the flaring $\gamma$-ray emission is associated with nonthermal
ions. Thus, the QPPs in $\gamma$-ray, HXR and radio/microwave
channels suggest the similar dynamical process during the impulsive
phase of a solar flare, i.e., nonthermal electrons and ions are
periodically accelerated by the magnetic energy released by e.g.,
repetitive magnetic reconnection \citep[see,][and references
therein]{Kliem00,Nakariakov09,Nakariakov18}. Moreover, the
cross-correlation coefficients between the QPP signals detected in
the G1 and G2 channels and the G2 and G3 channel are found to be
0.81 and 0.76, respectively. For all three channels the highest
correlation occurs for the zero time lags between the QPP signals,
indicating the cotemporal nature of the QPPs. However, it is
difficult to measure the correlation of the high-energy QPP signals
with the radio QPP signals, as the radio QPPs occur about 90~s later
than the HXR and $\gamma$-ray QPPs. But both the high-energy and
radio QPPs clearly show a similar oscillation period. On the other
hand, the observed periods of 20$-$30~s in the QPPs are very common
in solar flares, which are often explained as the sausage
oscillations at coronal or flaring loops
\citep[e.g.,][]{Inglis16,Tian16,McLaughlin18}. Then the slightly
different periods can be attributed to the gradual variations of the
physical parameters at the oscillation flaring loops, i.e., the
small variations of loop length or plasma density of flaring loops
\citep{Hayes16,Tian16,Kolotkov18}. However, we cannot conclude this,
because it is hard to detect these small-scale variations due to the
lack of high-resolution imaging observations. Noting that the AIA
images are saturated during the powerful solar flare.

\begin{figure}
\centering
\includegraphics[width=\linewidth,clip=]{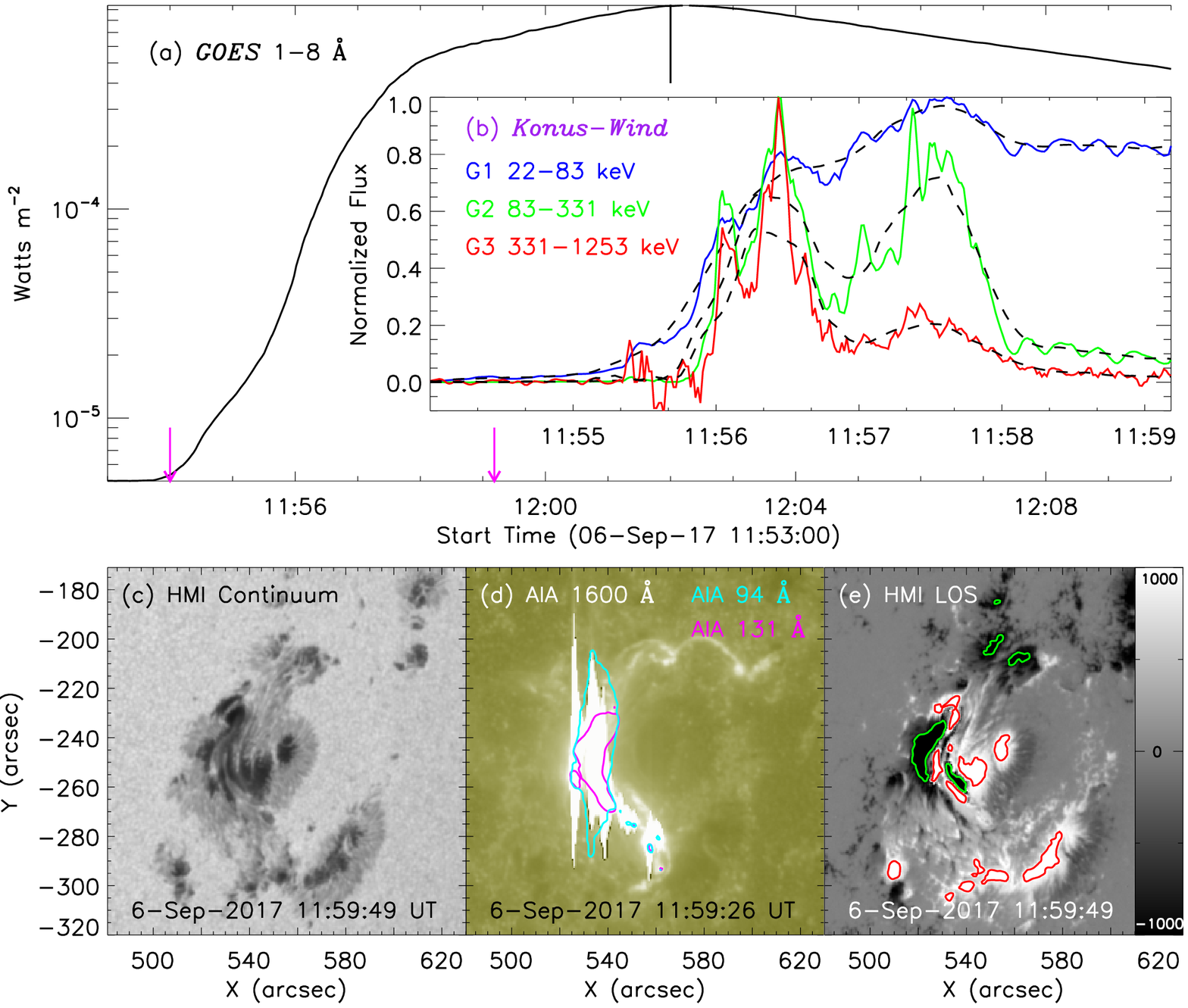}
\caption{Panels~(a) and (b): Light curves in {\it GOES} 1$-$8~{\AA}
(black), {\it Konus-Wind}~22$-$83~keV (blue), 83$-$331~keV (green),
and 331$-$1253~keV (red). The black dashed lines in panel~(b) show
the trended light curves. The short vertical line marks the flaring
peak time, and two magenta arrows outline the time range in
panel~(b). Panel~(c): HMI continuum filtergram. Panel~(d):
AIA~1600~{\AA} image, the magenta and cyan contours are derived from
AIA~131~{\AA} and 94~{\AA} images with levels of 20000~DN~s$^{-1}$
and 3000~DN~s$^{-1}$, respectively. Panel~(e): HMI LOS magnetogram,
the red and green contours represent the positive and negative
magnetic fields at the level of $\pm$1000~G. \label{image}}
\end{figure}

\begin{figure}
\centering
\includegraphics[width=\linewidth,clip=]{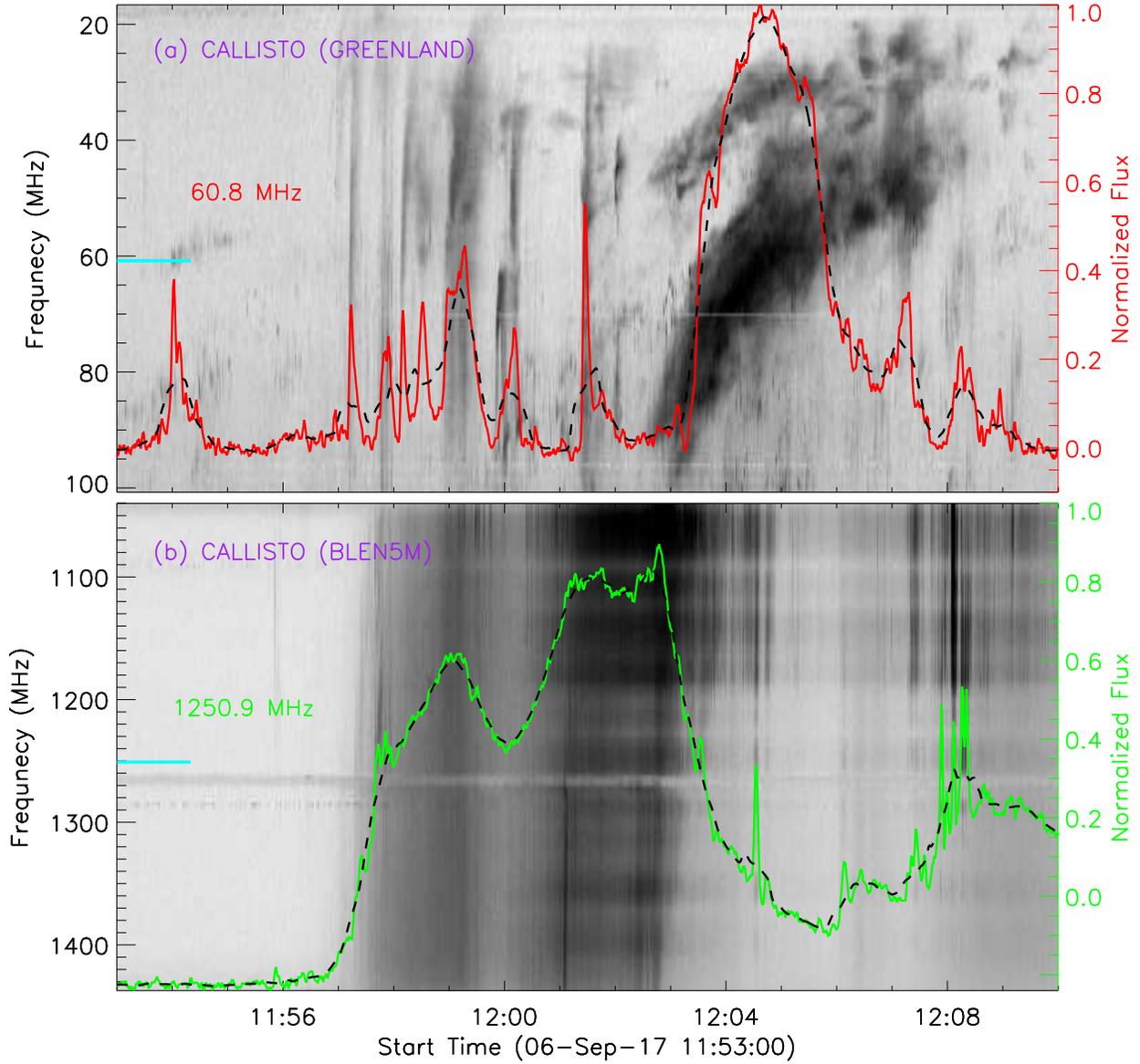}
\caption{Radio dynamic spectra recorded by GREENLAND (a), and BLEN5M
(b), respectively. The over-plotted light curves are the
radio/micowave fluxes indicated with a short cyan line on the
left-hand side of each image, and the black dashed lines show their
trended light curves. \label{spect}}
\end{figure}

\begin{figure}
\centering
\includegraphics[width=\linewidth,clip=]{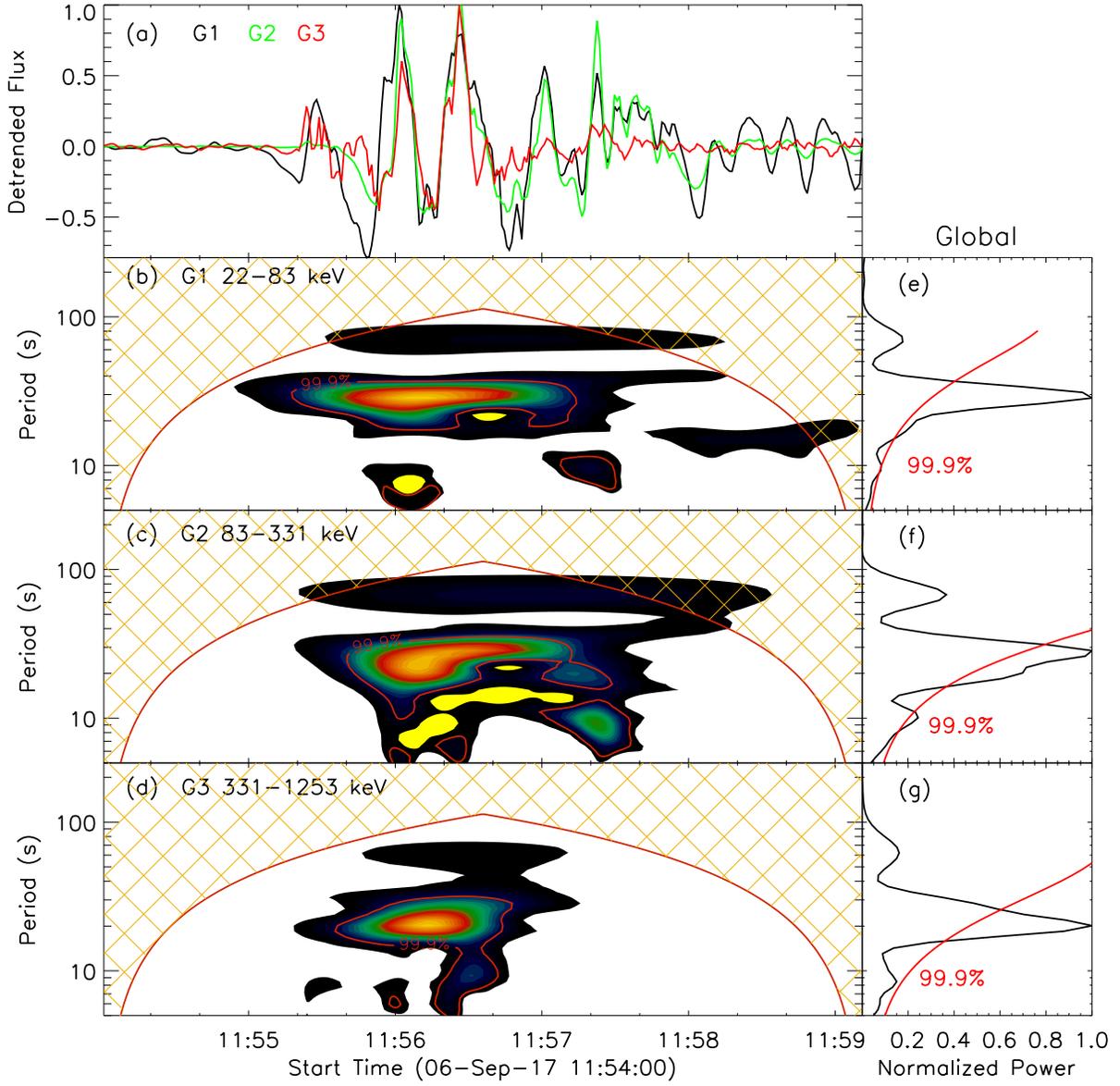}
\caption{Panel~(a): Normalized detrended light curves observed by
{\it Konus-Wind} in HXR and $\gamma$-ray channels. Panels~(b)$-$(d):
Their wavelet power spectra. Panels~(e)$-$(g): Their global wavelet
power. The red lines indicate a significance level of 99.9\%.
\label{wave1}}
\end{figure}

\begin{figure}
\centering
\includegraphics[width=\linewidth,clip=]{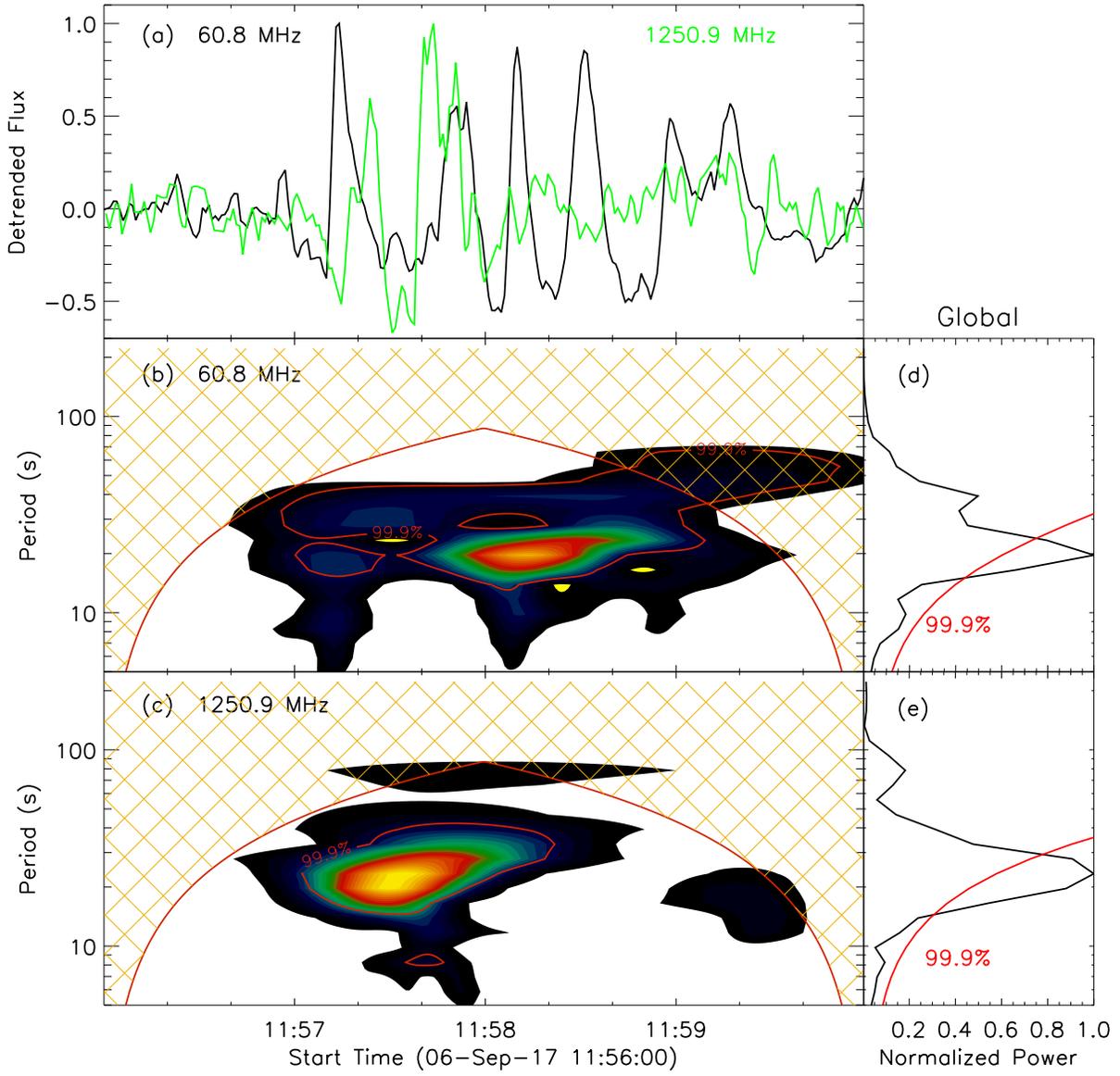}
\caption{Panel~(a): Normalized detrended light curves in
radio/macowave band. Panels~(b) and (c): Their wavelet power
spectra. Panels~(d) and (e): Their global wavelet power. The red
lines indicate a significance level of 99.9\%. \label{wave2}}
\end{figure}

\begin{figure}
\centering
\includegraphics[width=\linewidth,clip=]{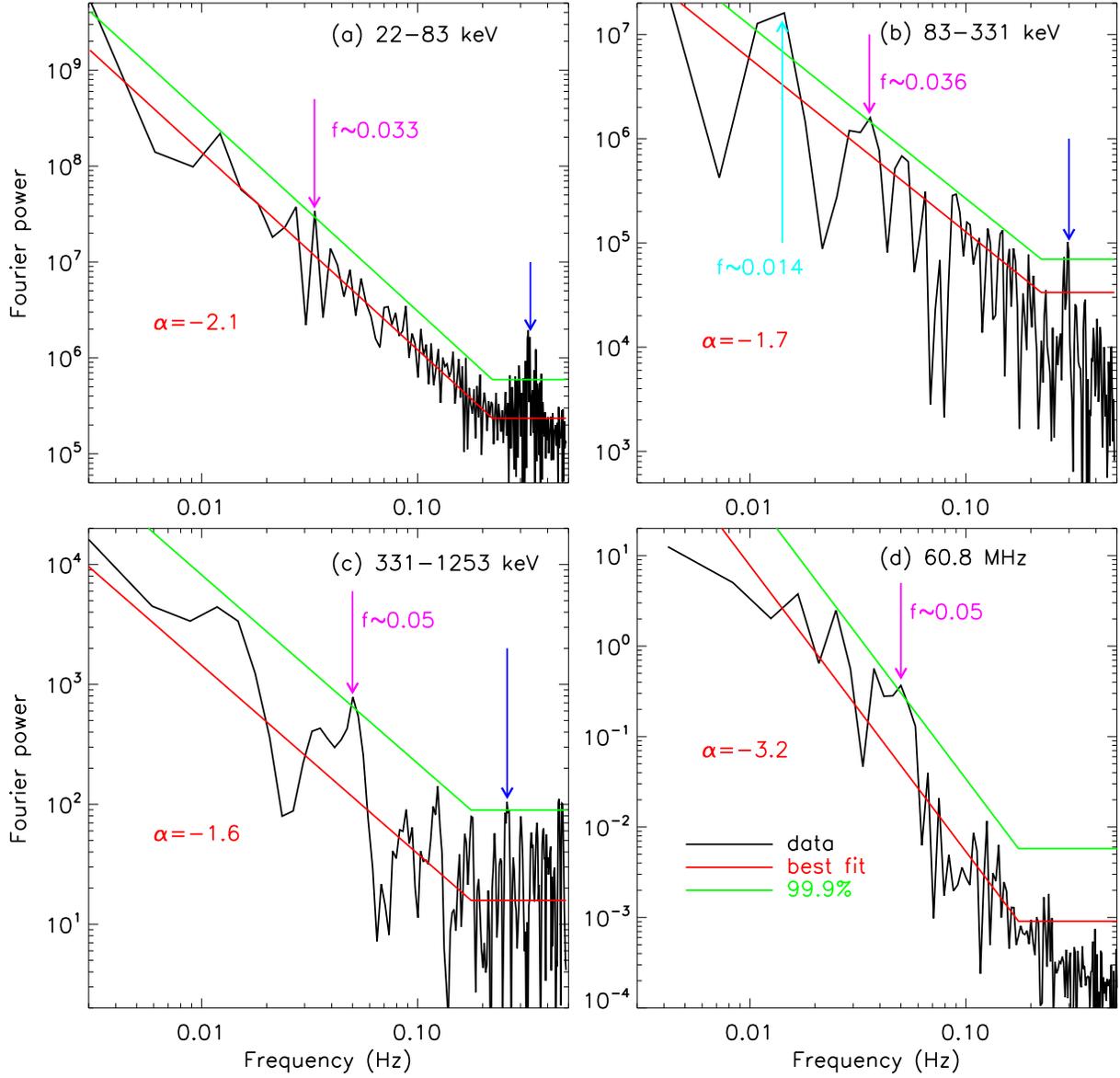}
\caption{Fourier power spectra of the $Konus-Wind$ light curves
(a-c) and radio flux (d). The red line shows the best fit result,
and the green line represents the confidence level at 99.9\%. The
color arrows indicate the frequencies above the confidence level.
\label{fftp}}
\end{figure}

It is also necessary to stress the time delay of QPPs can be found
in HXR, $\gamma$-ray and radio channels. The onset time of QPPs in
G1 channel of {\it Konus-Wind} is earlier ($\sim$20~s) than that in
G2 \& G3 channels. The time delay might be because that the X-ray
radiation in G1 channel contains some SXR emissions, since its
energy band is a little low, such as 22$-$83~keV. This is also
consistent with the fact that the flare itself appears in the
higher-energy bands later than in the lowest-energy one, as shown in
Figure~\ref{image}~(b). While the QPPs in G2 (HXR) and G3
($\gamma$-ray) channels appear nearly in the same time, further
confirming that they are caused by the accelerated charged particles
produced by the same process of magnetic reconnection. The lifetime
of QPPs in $\gamma$-rays is shorter than that in HXRs, which might
be due to that the ion is much heavier than electron, making it need
more energy to accelerate. Thus, the power becomes more localized in
time in the intensity wavelet from X-rays to $\gamma$-rays, as shown
in Figure~\ref{wave1}~(b)$-$(d). On the other hand, the lifetime of
QPPs in microwave emissions is close to that in $\gamma$-rays, but
shorter than that in HXRs, suggesting that it also needs a large
amount of energies to sustain radiation. The onset time of QPPs in
radio and microwave bands is delayed by $\sim$2~s, but they appear
much later than QPPs in HXR (G2) \& $\gamma$-ray (G3) bands, i.e.,
by nearly 90~s. The latter delay might be associated with the
radiation process (such as a specific velocity distribution)
produced radio/microwave emissions requiring a longer time than that
responsible for the radiation in HXR or $\gamma$-ray bands
\citep[see,][]{Nishizuka15}. An additional reason for the time delay
might be the time lag between reconnection and acceleration
processes, but it is still under discussions
\citep[e.g.,][]{Warmuth09}.

Finally, we notice that QPPs in the X9.3 flare have already been
reported by \cite{Kolotkov18}. Using the time derivatives of the
light curves measured by {\it GOES} and {\it SDO}/ESP, they found
the periods of QPPs drifting from 12~s to 25~s during the impulsive
and decay phases. The QPPs were interpreted as sausage oscillations
at flaring loops, but they could not rule out the other mechanisms,
i.e., repetitive reconnection. Time derivatives of SXR fluxes of
solar flares often closely match with the HXR/micorwave light
curves, which is known as the `Neupert effect'
\citep[e.g.,][]{Neupert68,Kahler70,Ning08}. However, the QPPs
detected in these wavebands in the analyzed flare are of different
periods, drifting from $\sim$12~s to $\sim$25~s
\citep[see,][]{Kolotkov18} and rather stable $\sim$28/24~s in our
study, respectively, suggesting different QPP-generation mechanisms
operating in those wavebands. The difference in the QPP periods by a
factor of two may also suggest that the oscillatory signals are
linked with each other by a square dependence, i.e. the oscillatory
signal detected by \cite{Kolotkov18} is a square of the signal
detected in this study. However, theoretical models predicting such
a dependence are absent. Also, in our analysis we did not find long
periods such as 4$-$5 minutes detected in this flare
\citep{Kolotkov18}, because the effective duration of light curves
recorded by {\it Konus-Wind} is too short to detect them. On the
other hand, the X9.3 flare was also observed by the Large-Yield
RAdiometer aboard the PROBA2, and their light curves showed a clear
signature of QPPs, as can be seen in Figure~2 of \cite{Dominique18}.
The QPPs during the impulsive phase described there are similar to
ours, but they are not discussed by the authors.

The QPPs of nonthermal emissions such as $\gamma$-ray, HXR, radio,
and microwave are detected in a powerful solar flare (X9.3), which
has an energy realm of the typical stellar flares
\citep{Kolotkov18}. So it is helpful to bridge the energy gap
between the solar and stellar flares \citep{Maehara15,Pugh16}, and
it is also interesting for investigating the flaring energy release
and particle acceleration on the Sun and Sun-like stars
\citep{Nakariakov10,McLaughlin18}.

\acknowledgments  We thank the anonymous referee for his/her
inspiring and valuable comments. The authors would like to
acknowledge Dr. S.~Anfinogentov for his inspiring discussion. We
thank the teams of {\it Konus-Wind}, {\it GOES}, {\it SDO}, and
CALLISTO (the Institute for Data Science, FHNW Brugg/Windisch,
Switzerland) for their open data use policy. This work is supported
by NSFC under grants 11973092, 11603077, 11573072,
11790300,11790302, 11729301, 11873095, the Youth Fund of Jiangsu
Nos. BK20161095, and BK20171108, as well as the Strategic Priority
Research Program on Space Science, CAS, Grant No. XDA15052200 and
XDA15320301. D.~Li is supported by the Specialized Research Fund for
State Key Laboratories. V.~M.~Nakariakov and D.~Y.~Kolotkov
acknowledge support by the STFC consolidated grant ST/P000320/1.
V.M. Nakariakov acknowledges the Russian Foundation for Basic
Research grant No. 17-52-80064 BRICS-A. The Laboratory No.
2010DP173032.

\end{document}